\documentclass[final,onefignum,onetabnum]{siamart250211}

\usepackage{braket,amsfonts}

\usepackage{array}

\usepackage[caption=false]{subfig}
\usepackage{pgfplots}

\newsiamthm{claim}{Claim}
\newsiamremark{remark}{Remark}
\newsiamremark{hypothesis}{Hypothesis}
\crefname{hypothesis}{Hypothesis}{Hypotheses}

\usepackage{algorithmic}

\usepackage{graphicx,epstopdf}

\Crefname{ALC@unique}{Line}{Lines}

\usepackage{amsopn}

\usepackage{xspace}
\usepackage{bold-extra}
\usepackage[most]{tcolorbox}

\colorlet{texcscolor}{blue!50!black}
\colorlet{texemcolor}{red!70!black}
\colorlet{texpreamble}{red!70!black}
\colorlet{codebackground}{black!25!white!25}

\lstdefinestyle{siamlatex}{%
  style=tcblatex,
  texcsstyle=*\color{texcscolor},
  texcsstyle=[2]\color{texemcolor},
  keywordstyle=[2]\color{texemcolor},
  moretexcs={cref,Cref,maketitle,mathcal,text,headers,email,url},
}

\tcbset{%
  colframe=black!75!white!75,
  coltitle=white,
  colback=codebackground, % bottom/left side
  colbacklower=white, % top/right side
  fonttitle=\bfseries,
  arc=0pt,outer arc=0pt,
  top=1pt,bottom=1pt,left=1mm,right=1mm,middle=1mm,boxsep=1mm,
  leftrule=0.3mm,rightrule=0.3mm,toprule=0.3mm,bottomrule=0.3mm,
  listing options={style=siamlatex}
}

\newtcblisting[use counter=example]{example}[2][]{%
  title={Example~\thetcbcounter: #2},#1}

\newtcbinputlisting[use counter=example]{\examplefile}[3][]{%
  title={Example~\thetcbcounter: #2},listing file={#3},#1}

\DeclareTotalTCBox{\code}{ v O{} }
{ %fontupper=\ttfamily\color{texemcolor},
  fontupper=\ttfamily\color{black},
  nobeforeafter,
  tcbox raise base,
  colback=codebackground,colframe=white,
  top=0pt,bottom=0pt,left=0mm,right=0mm,
  leftrule=0pt,rightrule=0pt,toprule=0mm,bottomrule=0mm,
  boxsep=0.5mm,
  #2}{#1}

\patchcmd\newpage{\vfil}{}{}{}
\begin{tcbverbatimwrite}{tmp_\jobname_header.tex}
\title{Arginine metabolism drives tumor immune escape by shaping an immunosuppressive microenvironment \thanks{Received by the editors XXXX, XXXX; accepted for publication (in revised form) XXXX, XXXX; published electronically XXXX, XXXX. \funding{This work was funded by the Key Program of National Natural Science Foundation of China (NSFC 12331018).}} }

\author{
	Chenghang Li\thanks{School of Mathematical Sciences, Tiangong University, Tianjin, 300387, People's Republic of China (\email{lichenghang@tiangong.edu.cn, quyipu@tiangong.edu.cn}).} 
	\and Haifeng Zhang\thanks{School of Mathematical Sciences, Jiangsu University, Zhenjiang, 212013, People's Republic of China (\email{haifengzhang202212@163.com}).}
	\and Yipu Qu$^{\dag}$
	\and Jinzhi Lei$^{\dag}$\thanks{Corresponding author. School of Mathematical Sciences, Center for Applied Mathematics, Tiangong University, Tianjin, 300387, People's Republic of China (\email{jzlei@tiangong.edu.cn}).}
}

\headers{Mathematical model of the immunosuppressive mechanism}{C. LI, H. ZHANG, Y. QU, AND J. LEI}
\end{tcbverbatimwrite}
\input{tmp_\jobname_header.tex}

\ifpdf
\hypersetup{ pdftitle={Guide to Using  SIAM'S \LaTeX\ Style} }
\fi

\begin{document}
\maketitle

%% ------------------------------------------------------------------
%% ABSTRACT
%% ------------------------------------------------------------------
\begin{tcbverbatimwrite}{tmp_\jobname_abstract.tex}
\begin{abstract}
The tumor microenvironment evolves through the ``elimination-equilibrium-escape'' paradigm, yet the quantitative mechanisms by which metabolic reprogramming contributes to immune evasion remain unclear. We develop a mathematical model of tumor-immune interactions to investigate how arginine metabolism shapes an immunosuppressive microenvironment. We establish the positivity and boundedness of solutions and analyze the existence and stability of equilibrium states. The model is calibrated using tumor growth data from three mouse cohorts and reproduces tumor growth patterns under distinct immune conditions. The calibrated model predicts faster tumor growth in immunocompetent mice than in immunodeficient controls, with macrophage dynamics accounting for this difference. It further predicts that arginine administration promotes tumor progression by enhancing macrophage polarization and weakening T cell-mediated tumor control. To account for interindividual heterogeneity, we develop the TVInfABC algorithm to generate a virtual mouse cohort. Moreover, our analysis indicates that tumor-associated macrophages can promote transitions from immunologically ``hot'' to ``cold'' states. Systematic pharmacodynamic simulations further uncover a synergy between arginine depletion (anti-ARG1) and immune checkpoint blockade (anti-PD-1/PD-L1). These results provide a mathematical basis for studying metabolic-immune interactions and for evaluating mechanism-based combination therapies.
\end{abstract}



\end{tcbverbatimwrite}
\input{tmp_\jobname_abstract.tex}
%% ------------------------------------------------------------------
%% END HEADER
%% ------------------------------------------------------------------

\section{Introduction}

Tumor progression is not determined by tumor cell proliferation alone, but also by the evolving balance between antitumor immunity and immunosuppressive regulation in the tumor microenvironment \cite{Joyce.Science.2015}. This balance underlies cancer immunoediting, in which tumors may be eliminated, held in an equilibrium state, or eventually escape immune control \cite{Schreiber.Science.2011}. A central unresolved question is how metabolic changes in the tumor microenvironment alter this balance and promote the transition to immune escape. Mathematical models of tumor-immune dynamics have often represented tumor cells and T cells through a predator-prey interaction \cite{Dehingia.MathComputSimulat.2023,Han.ApplMathModel.2022}. Although such models describe direct immune-mediated tumor killing, they do not explicitly account for immunosuppressive cells that can reshape T cell activity and alter the long-term behavior of the system. Tumor-associated macrophages are particularly relevant in this regard, as they can be recruited by tumors, acquire immunosuppressive phenotypes, and weaken T cell-mediated tumor control. In this work, we develop a mathematical model for the interactions among tumor cells, tumor-specific T cells, and tumor-associated macrophages. The model incorporates arginine-mediated macrophage polarization to examine how metabolic reprogramming can reshape the tumor microenvironment and facilitate immune escape.

The coupled interactions among tumor cells, immune effector cells, and immunosuppressive cells can generate nonlinear dynamics across multiple time scales. Therefore, nonlinear dynamical systems provide a natural framework for examining how these interactions determine tumor control, immune equilibrium, and immune escape. Mathematical models provide quantitative tools for investigating tumor-immune interactions across multiple biological scales \cite{Eftimie.BullMathBiol.2016,Li.CSIAM-LS.2025,Sun.BriefBioinform.2018}. Early studies showed that even relatively simple tumor-immune models can exhibit rich nonlinear dynamics. In particular, bifurcation analyses have identified Bogdanov-Takens bifurcations in vascularized and avascular tumor settings \cite{Liu.DCDSB.2009}, while two-stage tumor models have revealed stable periodic solutions associated with alternating phases of immune surveillance and tumor escape \cite{Liu.MathBiosciEng.2012}. These results demonstrate that tumor progression and immune control may depend on both parameter regimes and dynamical states.

More recent models have incorporated additional components of the tumor microenvironment to examine immune suppression and therapeutic responses. For example, ordinary differential equation models combined with experimental data and approximate Bayesian computation have been used to study the effects of immune checkpoints and myeloid-derived suppressor cells on T cell activity \cite{Anderson.JMathBiol.2023}. Tumor ecosystem models have further considered the coupled roles of tumor, immune, and mutualistic cell populations in cancer progression \cite{Wang.SIAMJApplMath.2025}, whereas models of oncolytic virotherapy have examined the interactions between oncolytic viruses and cytotoxic T lymphocytes \cite{Zhao.SIAMJApplMath.2025}. At broader spatial and biological scales, multiscale models have been developed to investigate macrophage-mediated resistance to immunotherapy \cite{Lin.MultiscaleModelSim.2025}, and nonlocal reaction-diffusion models informed by spatial transcriptomic data have been used to study tumor phenotypic heterogeneity \cite{Li.JMathBiol.2026}.

Together, these studies show that immunosuppressive cells, therapeutic perturbations, and tumor heterogeneity can substantially affect tumor-immune dynamics. However, the dynamical role of tumor-associated macrophages in linking metabolic reprogramming to T cell suppression and immune escape remains insufficiently understood. In particular, it is unclear how arginine-mediated macrophage polarization changes the equilibrium structure and long-term dynamics of the tumor-immune system. The model developed here addresses this question by incorporating tumor cells, tumor-specific T cells, and tumor-associated macrophages into a mechanistic ordinary differential equation framework.

Many biological processes involved in tumor-immune interactions, including immune cell activation, recruitment, and treatment response, may involve non-negligible time lags. Delay differential equations therefore provide a useful framework for studying biological systems in which the current dynamics depend on past states \cite{Martin.JMathBiol.2001,Ruan.MathModelNatPheno.2009}. In tumor-immune models, such delays can substantially alter equilibrium stability and generate qualitatively different dynamical outcomes \cite{Ruan.DCDSB.2021}. For example, models with multiple delays in tumor-effector cell interactions have exhibited periodic solutions and chaotic dynamics under different delay regimes \cite{Bi.SIAMJApplDynSyst.2013,Bi.Chaos.2014}. 

Time-delay models have also been used to study treatment-related tumor-immune dynamics. Multi-compartment delay systems have been developed to investigate the effects of dendritic cell vaccines on tumor control, elimination, and escape \cite{Dickman.SIAMJApplMath.2020}. In related work, bifurcation analysis of models with dendritic cell therapy and delayed immune responses has identified parameter thresholds associated with tumor escape and tumor-T cell coexistence \cite{Zhang.SIAMJApplMath.2023}. Moreover, the inclusion of immune checkpoint inhibition in a delayed tumor-immune model can modify system stability and induce oscillatory behavior through Hopf bifurcation \cite{Shi.CNSNS.2023}. These studies show that temporal delays can be important determinants of tumor-immune dynamics and treatment outcomes.

In the present study, we focus on the effects of arginine-mediated macrophage polarization on the equilibrium structure and long-term dynamics of the tumor-immune system. Here, we explore a biological mechanism through which tumors may establish an immunosuppressive microenvironment. Although tumor-T cell interactions are central to antitumor immunity \cite{Chen.Immunity.2013}, tumor-associated macrophages can substantially modify the effectiveness of T cell-mediated tumor control \cite{Lin.MultiscaleModelSim.2025,Liu.SciAdv.2025,Ni.SciAdv.2022,Zheng.MolCancerTher.2018}. Tumor cells can recruit macrophages into tumor tissues through the secretion of the chemokine CCL2 \cite{Cassetta.NatRevCancer.2023}. In addition, resting macrophages can take up tumor-derived arginine and convert it into polyamines, a process that promotes macrophage polarization toward an immunosuppressive phenotype \cite{Zhu.CancerCell.2025}. These polarized tumor-associated macrophages can subsequently suppress tumor-specific T cell responses \cite{Kloosterman.Cell.2023}. This sequence of tumor-driven recruitment, arginine-mediated polarization, and T cell suppression provides a mechanistic basis for studying how metabolic reprogramming contributes to immune escape.

Motivated by these observations, we develop an ordinary differential equation model for the interactions among tumor cells, tumor-specific T cells, and tumor-associated macrophages. The model incorporates CCL2-mediated macrophage recruitment, arginine-mediated macrophage polarization, and macrophage-induced suppression of T cell activity. We first establish the positivity and boundedness of solutions and analyze the existence and stability of equilibrium states. We then calibrate the model using experimental tumor growth data and perform global sensitivity analysis. To characterize interindividual variability, we develop the Tumor Volume Inference using Approximate Bayesian Computation (TVInfABC) method to generate a virtual cohort. This cohort is used to evaluate 15 combination treatment strategies. We further investigate the conditions under which the tumor-immune system exhibits bistable dynamics through bifurcation analysis. Our analysis indicates that macrophage polarization can promote the transition from an immunologically ``hot'' tumor state to a ``cold'' tumor state. It further suggests that ARG1 inhibition combined with immune checkpoint blockade can reduce tumor burden and produce a synergistic effect in the model.

The remainder of this paper is organized as follows. Section 2 formulates the tumor-immune ordinary differential equation model and describes the underlying biological assumptions. Section 3 establishes the main mathematical properties of the model. Section 4 calibrates the model with experimental data, constructs a virtual cohort using approximate Bayesian computation, and evaluates combination treatment strategies. Section 5 discusses the main findings, limitations, and directions for future work.

\section{Mathematical model}

\begin{figure}[h!]
	\centering
	\includegraphics[width=12.5cm]{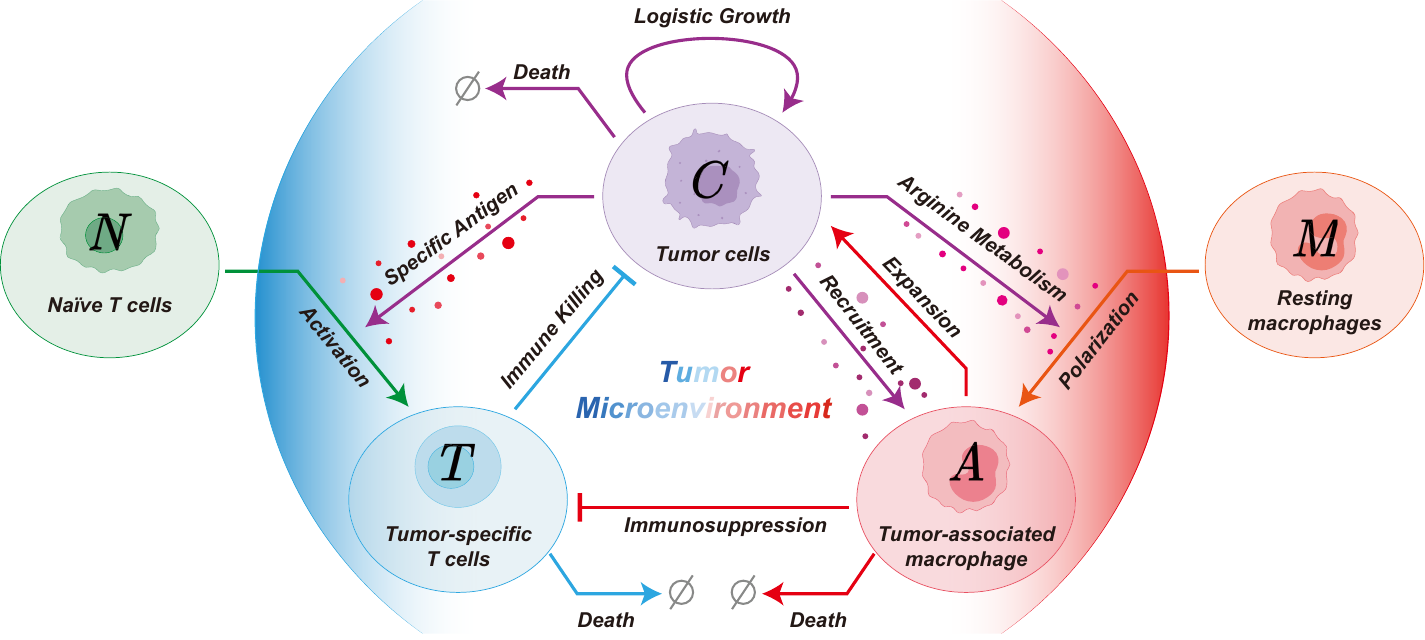}
	\caption{Regulatory network of tumor-immune interactions. The model includes three dynamic variables: tumor cells ($C$), tumor-specific T cells ($T$), and tumor-associated macrophages ($A$), together with two constant cell populations: na\"{i}ve T cells ($N$) and resting macrophages ($M$).}
	\label{Fig1}
\end{figure}

We consider a three-dimensional ordinary differential equation model of interactions among tumor cells, tumor-specific T cells, and tumor-associated macrophages. The regulatory pathways are summarized in Fig.~\ref{Fig1}.

\textbf{Equation for tumor cells ($C$).} We assume that tumor cells ($C$) undergo logistic growth. Tumor-specific T cells ($T$) are primary mediators of tumor killing \cite{Chen.Immunity.2013}, whereas tumor-associated macrophages ($A$) can promote tumor proliferation and angiogenesis \cite{Kloosterman.Cell.2023}. Accordingly, the dynamics of tumor cells, including proliferation, immune-mediated killing, and natural death, are described by
\begin{eqnarray}
	\frac{\mathrm{d} C}{\mathrm{d} t} = \underbrace{\alpha (1+\epsilon A) C \left( 1 -\frac{C}{\beta}\right)}_{\rm Logistic\ Growth} - \underbrace{\eta T C}_{\rm Immune\ Killing} - \underbrace{d_C C}_{\rm Death},
\end{eqnarray}
where $\alpha$ is the intrinsic tumor proliferation rate, $\epsilon$ is the coefficient quantifying macrophage-mediated tumor growth promotion, $\beta$ is the tumor carrying capacity, $\eta$ is the T cell-mediated killing rate of tumor cells, and $d_C$ is the natural death rate.

\textbf{Equation for tumor-specific T cells ($T$).} Na\"{i}ve T cells ($N$) are activated into tumor-specific T cells ($T$) upon recognition of tumor-specific antigens \cite{Chen.Immunity.2013}. Tumor-associated macrophages ($A$) suppress T cell activity through multiple immunosuppressive mechanisms \cite{Cassetta.NatRevCancer.2023,Kloosterman.Cell.2023,Zhu.CancerCell.2025}. Accordingly, the dynamics of tumor-specific T cells are described by 
\begin{eqnarray}
\frac{\mathrm{d} T}{\mathrm{d} t} = \underbrace{\kappa \frac{C}{\omega+C} N}_{\rm Immune\ Activation} - \underbrace{\delta A T}_{\rm Immunosuppression}-\underbrace{d_T T}_{\rm Death}, 
\end{eqnarray}
where $\kappa$ denotes the maximal activation rate, $\omega$ is the half-saturation constant for tumor antigen-dependent activation, $\delta$ is the macrophage-mediated suppression rate, and $d_T$ is the natural death rate.

\textbf{Equation for tumor-associated macrophages ($A$).} As the most abundant innate immune cells, tumor-associated macrophages ($A$) constitute the predominant cellular component of the immunosuppressive tumor microenvironment \cite{Cassetta.NatRevCancer.2023,Kloosterman.Cell.2023,Zhu.CancerCell.2025}. In our model, $A$ arises through two mechanisms: recruitment of macrophages by tumor-derived CCL2 \cite{Cassetta.NatRevCancer.2023} and polarization of resting macrophages ($M$) toward a pro-tumor phenotype by tumor-derived arginine \cite{Zhu.CancerCell.2025}. Therefore, the dynamics of tumor-associated macrophages, including recruitment, polarization, and death, are described by 
\begin{eqnarray}
\frac{\mathrm{d} A}{\mathrm{d} t} = \underbrace{\chi \frac{C}{\sigma+C}}_{\rm Recruitment}+ \underbrace{\gamma \frac{C}{\theta + C} M}_{\rm Polarization} - \underbrace{d_{A} A}_{\rm Death},
\end{eqnarray}
where $\chi$ is the maximum recruitment rate of macrophages induced by tumor-derived CCL2, $\sigma$ is the corresponding half-saturation constant, $\gamma$ is the maximum rate of arginine-mediated macrophage polarization, $\theta$ is the corresponding half-saturation constant, and $d_{A}$ is the natural death rate. Within the tumor microenvironment, resting macrophages ($M$) are continuously replenished through the infiltration of blood-borne monocytes. We assume that this replenishment compensates for the loss of resting macrophages due to polarization on the time scale considered. Hence, $M$ is treated as a constant external cell source.

\section{Theoretical analysis}

Based on the model formulation above, the interactions among tumor cells ($C$), tumor-specific T cells ($T$), and tumor-associated macrophages ($A$) are governed by
\begin{equation}
	\label{eq:model}
	\left\{
	\begin{aligned}
		\frac{\mathrm{d} C}{\mathrm{d} t} &= \underbrace{\alpha (1+ \epsilon A) C \left( 1 -\frac{C}{\beta}\right)}_{\rm Logistic\ Growth} - \underbrace{\eta T C}_{\rm Immune\ Killing} - \underbrace{d_C C}_{\rm Death},\\
		\frac{\mathrm{d} T}{\mathrm{d} t} &= \underbrace{\kappa \frac{C}{\omega+C} N }_{\rm Immune\ Activation} - \underbrace{\delta A T}_{\rm Immunosuppression}-\underbrace{d_T  T}_{\rm Death}, \\
		\frac{\mathrm{d} A}{\mathrm{d} t} &= \underbrace{ \chi  \frac{C}{\sigma+C} }_{\rm Recruitment} + \underbrace{\gamma \frac{C}{\theta + C} M}_{\rm Polarization} - \underbrace{d_{A} A}_{\rm Death}. 
	\end{aligned}
	\right. 
\end{equation}
All parameters and the constant cell populations $N$ and $M$ are assumed to be positive. In the following, we investigate the qualitative properties of system \eqref{eq:model}.

\subsection{Positivity and boundedness} 

We first establish the positivity and boundedness of solutions to ensure the biological feasibility of the model. In particular, the cell populations remain nonnegative and cannot grow without bound.

\begin{theorem}[Positivity and boundedness]
\label{thm:well_posedness}
Let the initial conditions satisfy $C(0) > 0$, $T(0) > 0$, and $A(0) > 0$. Then, system \eqref{eq:model} admits a unique solution $(C(t), T(t), A(t))$ for all $t \geq 0$, and this solution remains positive and bounded. More precisely, 
\begin{equation}
0 < C(t)\leq \hat{C},\quad 0 < T(t)\leq \hat{T}, \quad 0 < A(t)\leq \hat{A}, \quad t\geq 0.
\end{equation}
Here, $\hat{A} = \max \left\{A(0), \frac{\chi+\gamma M}{d_A}\right\}$, $\hat{T} = \max \left\{T(0), \frac{\kappa N}{d_T}\right\}$, and $\hat{C}  = \max \left\{C(0), \frac{\alpha \beta (1+ \epsilon \hat{A})}{4 d_C}\right\}$.
\end{theorem}

\begin{proof}
Since the right-hand side of system \eqref{eq:model} is locally Lipschitz continuous in $\mathbb{R}_{>0}^3$, the system admits a unique local solution.

\textit{Positivity:} The first equation of system \eqref{eq:model} yields
\begin{equation}
C(t) = C(0) \exp \left( \int_0^t \left[ \alpha (1+ \epsilon A(s) ) \left( 1 -\frac{C(s) }{\beta}\right) - \eta T(s) - d_C \right] \mathrm{d} s \right).
\end{equation}
Thus, $C(t)>0$ for all times for which the solution exists. Moreover, the third equation gives 
$$\frac{\mathrm{d} A}{\mathrm{d} t} \geq -d_A A,$$ 
and hence $A(t) \geq A(0) e^{-d_A t} > 0$. Finally, since $C(t)> 0$ and $A(t) > 0$, 
$$\frac{\mathrm{d} T}{\mathrm{d} t} \geq -\left(\delta A(t) + d_T \right) T,$$ 
which implies $T(t) > 0$.

\textit{Boundedness:} Since $0\leq \frac{C}{\sigma + C} < 1$ and $0\leq \frac{C}{\theta + C} < 1$, the third equation of system \eqref{eq:model} satisfies 
$$\frac{\mathrm{d} A}{\mathrm{d} t} \leq \chi + \gamma M - d_A A.$$ 
The comparison principle then gives
\begin{equation}
A(t) \leq \max\left\{A(0), \frac{\chi + \gamma M}{d_A}\right\} = \hat{A}.
\end{equation}
Similarly, 
$$\frac{\mathrm{d} T}{\mathrm{d} t} \leq \kappa N - d_T T,$$ 
and thus
\begin{equation}
T(t) \leq \max\left\{T(0), \frac{\kappa N}{d_T}\right\} = \hat{T}.
\end{equation}
Finally, for $C\geq 0$, the function $g(C) = C(1 - C/\beta)$ satisfies $g(C) \leq \frac{\beta}{4}$. Since $A(t) \leq \hat{A}$ for all $t \geq 0$, the first equation of system \eqref{eq:model} implies 
$$\frac{\mathrm{d} C}{\mathrm{d} t} \leq \frac{\alpha\beta (1 + \epsilon\hat{A})}{4} -d_C C.$$  
Applying the comparison principle, we obtain
\begin{equation}
	C(t) \le \max\left\{C(0), \frac{\alpha\beta(1 + \epsilon \hat{A})}{4 d_C} \right\} = \hat{C}.
\end{equation}
Hence, the local solution remains in a bounded subset of $\mathbb{R}_{>0}^3$ and can be extended to all $t\geq 0$. This completes the proof.
\end{proof}

\begin{remark}
\label{rem:3.2}
As a direct consequence of Theorem \ref{thm:well_posedness}, let
$$\bar{A}=\frac{\chi+\gamma M}{d_A},\qquad \bar{T}=\frac{\kappa N}{d_T},\qquad \bar{C}=\frac{\alpha\beta(1+\epsilon\bar{A})}{4d_C},$$ 
the set
\begin{equation}
\Omega=
\left\{
(C, T, A)\in\mathbb{R}_{\geq0}^3:
0\leq C\leq \bar{C},\,
0\leq T\leq \bar{T},\,
0\leq A\leq \bar{A}
\right\},
\end{equation}
is positively invariant for system \eqref{eq:model}. Indeed, if
$(C(0),T(0),A(0))\in\Omega$, Theorem \ref{thm:well_posedness}
implies that $(C(t),T(t),A(t))\in\Omega$, $t\geq0$.
\end{remark}

\subsection{Equilibrium structure of the model system}

With the positivity and boundedness of solutions established, we next examine the equilibrium structure of system \eqref{eq:model}. The system always admits the tumor-free equilibrium $E_0 = (0, 0, 0)$, which corresponds to tumor eradication. We first investigate its local stability.

\begin{theorem}[Local stability of the tumor-free equilibrium]
\label{thm:TFE_local}
The tumor-free equilibrium $E_0 = (0,0,0)$ is locally asymptotically stable if $\alpha < d_C$ and unstable if $\alpha > d_C$.
\end{theorem}

\begin{proof}
The Jacobian matrix of system \eqref{eq:model} at $E_0$ is \begin{equation}
\label{eq:j2}
J(E_0) =  
\begin{pmatrix}
\alpha  -d_C   & 0   & 0    \\
\dfrac{\kappa N}{\omega}  & -d_T   & 0    \\
\dfrac{ \chi }{ \sigma } + \dfrac{\gamma M}{ \theta}  & 0   & -d_{A}         
\end{pmatrix}.
\end{equation}
Since $J(E_0)$ is lower triangular, its eigenvalues are
\begin{equation}
\lambda_1 = \alpha - d_C, \ \lambda_2 = -d_T, \ \lambda_3 = -d_A. 
\end{equation}
Because $d_T > 0$ and $d_A > 0$, all eigenvalues have negative real parts when $\alpha < d_C$. Therefore, $E_0$ is locally asymptotically stable in this case. If $\alpha > d_C$, then $\lambda_1 > 0$, and hence $E_0$ is unstable.
\end{proof}

The case $\alpha = d_C$ is nonhyperbolic and is not resolved by linearization alone. We next derive a sufficient condition for global tumor elimination. From the third equation of system \eqref{eq:model}, 
$$\frac{\mathrm{d} A}{\mathrm{d} t} \leq \chi + \gamma M - d_A A,$$ which implies $\limsup_{t\to\infty}A(t) \leq \frac{\chi + \gamma M}{d_A} = \bar{A}$. Using this asymptotic upper bound, the following theorem provides a sufficient condition for global convergence to the tumor-free equilibrium.

\begin{theorem}[Global asymptotic stability of the tumor-free equilibrium]
\label{thm:TFE_global}
If $\alpha (1+ \epsilon \bar{A}) < d_C$, then the tumor-free equilibrium $E_0 = (0,0,0)$ is globally asymptotically stable in $\Omega$.
\end{theorem}

\begin{proof}
By Remark \ref{rem:3.2}, $\Omega$ is positively invariant for system \eqref{eq:model}. Hence, it suffices to show that every solution starting in $\Omega$ converges to $E_0$. Since $\limsup_{t \to \infty} A(t) \leq \bar{A}$, for any sufficiently small $\xi > 0$, there exists $t_1 > 0$ such that $A(t) < \bar{A} + \xi$ for all $t \geq t_1$. Choose $\xi$ such that
\begin{equation}
\lambda_\xi:= \alpha\bigl(1+\epsilon(\bar{A} + \xi)\bigr) - d_C < 0.
\end{equation} 
Then, for $t\geq t_1$, $\frac{\mathrm{d} C}{\mathrm{d} t} \leq \lambda_{\xi} C$, and hence $C(t) \leq C(t_1) e^{\lambda_{\xi}(t-t_1)}$. Therefore, $C(t)\to 0$ exponentially as $t\to\infty$.

As $C(t) \to 0$, $\kappa \frac{C(t)}{\omega + C(t)} N \to 0$ and $\chi\frac{C(t)}{\sigma+C(t)} + \gamma  \frac{C(t)}{\theta+C(t)} M \to 0$. Moreover, 
\begin{equation}
\frac{\mathrm{d} T}{\mathrm{d} t} \leq \kappa \frac{C(t)}{\omega + C(t)} N - d_T T,\quad \frac{\mathrm{d} A}{\mathrm{d} t} = \chi \frac{C}{\sigma + C} + \gamma M \frac{C}{\theta + C} - d_A A.
\end{equation}
By the comparison principle, it follows that $T(t) \to 0$ and $A(t) \to 0$ as $t\to\infty$. Hence, every solution starting in $\Omega$ converges to $E_0$.

Finally, the assumed condition implies $\alpha < d_C$. Thus, $E_0$ is locally asymptotically stable by Theorem \ref{thm:TFE_local}, which completes the proof.
\end{proof}

Numerical illustrations supporting the stability of the tumor-free equilibrium are provided in SM1.1.

When $\alpha > d_C$, the tumor-free equilibrium is unstable. We therefore consider the existence and uniqueness of a positive equilibrium $E^* = (C^*, T^*, A^*)$.

\begin{theorem}[Existence and uniqueness of the positive equilibrium]
	\label{thm:PE_existence}
	If $\alpha > d_C$, system \eqref{eq:model} admits at least one positive equilibrium. Furthermore, the positive equilibrium is unique if the following conditions hold:
	\begin{equation}
		\label{eq:epe1}
		\omega \geq \max \{\theta, \sigma \}
	\end{equation}
	and
	\begin{equation}
		\label{eq:epe2}
		\dfrac{\epsilon}{d_A} \Big( \dfrac{\chi}{\sigma}  + \dfrac{\gamma M}{\theta} \Big) \leq \dfrac{1}{\beta}.
	\end{equation}
\end{theorem}

\begin{proof}
		For $C \geq 0$, define
		\begin{equation}
			A(C) = \frac{1}{d_A}\left(\chi \frac{C}{\sigma + C} + \gamma M \frac{C}{\theta+C} \right)
		\end{equation}
		and
		\begin{equation}
			\label{eq:f12}
			\begin{aligned}
				f_1(C) &= \frac{1}{\eta}\left[\alpha \bigl(1 + \epsilon A(C) \bigr) \left(1 - \frac{C}{\beta}\right)  - d_C \right],\\
				f_2(C) &= \frac{\kappa NC}{(\omega + C) \bigl(d_T + \delta A(C)\bigr)}.
			\end{aligned}
		\end{equation}
		At a positive equilibrium, $A^* = A(C^*)$ and $T^* = f_1(C^*) = f_2(C^*)$. Hence, it suffices to solve
		$f(C):=f_1(C) - f_2(C) = 0$.
	
	Since $\alpha > d_C$, we have $f(0) = (\alpha - d_C)/\eta > 0$. Moreover, $f_1(\beta)=-d_C/\eta<0$ and $f_2(\beta)>0$, so $f(\beta) < 0$. The continuity of $f$ therefore yields a root $C^* \in (0, \beta)$. Since $f_2(C^*) > 0$, it follows that $T^* = f_1(C^*) = f_2(C^*) > 0$, while $A^* = A(C^*) > 0$. Thus, a positive equilibrium exists.
	
	It remains to establish uniqueness. Condition \eqref{eq:epe2} gives
	\begin{equation}
		\label{eq:re_ap3}
		\begin{aligned}
				f_1'(C) =& \dfrac{1}{\eta } \Big(-\dfrac{1}{\beta} \Big) \left( \alpha \left( 1+  \dfrac{\epsilon }{ d_A } \left(  \chi  \frac{C}{\sigma+C}  + \gamma  \frac{C}{\theta + C} M  \right)  \right)   \right)  \\
			&+  \dfrac{1}{\eta } \Big(1 -\dfrac{C}{\beta} \Big) \alpha  \dfrac{\epsilon }{ d_A } \left(  \chi  \frac{\sigma}{ (\sigma+C)^2 }  + \gamma  \frac{\theta}{ (\theta + C)^2 } M  \right)   \\ 
			\leq &  - \frac{\alpha }{\eta \beta} +  \frac{\alpha \epsilon}{\eta d_A}   \left(\frac{\chi \sigma}{ (\sigma+C)^2 }  + \frac{\gamma M \theta}{ (\theta + C)^2 }  \right)   \\
			< & \frac{\alpha }{\eta}\left[\frac{\epsilon }{ d_A } \left(\frac{\chi}{ \sigma }  + \frac{\gamma M}{ \theta } \right)  - \dfrac{1}{\beta}   \right] \leq 0
		\end{aligned}
	\end{equation}
	for all $C \in (0, \beta)$. Hence, $f_1'(C) < 0$.
	
	Furthermore, 
	\begin{equation}
		\label{eq:re_ap4}
		f_2'(C) = \frac{\kappa N \left[\omega d_T + \delta \bigl(\omega A(C) - C(\omega + C) A'(C) \bigr)\right]}{(\omega + C)^2\bigl(d_T + \delta A(C)\bigr)^2}.
	\end{equation}
	A direct calculation gives
	\begin{equation}
		\label{eq:re_ap5}
		\begin{aligned}
			\omega A(C)-C(\omega+C) A^{\prime}(C)=\frac{C^2}{d_A}\left[\frac{\chi(\omega-\sigma)}{(\sigma+C)^2}+\frac{\gamma M(\omega-\theta)}{(\theta+C)^2}\right] .
		\end{aligned}
	\end{equation}
	Under condition \eqref{eq:epe1}, the right-hand side is nonnegative. Thus, $f_2'(C)>0$ for $C>0$, and consequently $f'(C) < 0$ on $(0, \beta)$. Therefore, $f(C) = 0$ has a unique root, which proves the uniqueness of the positive equilibrium.
\end{proof}

Numerical illustrations of the existence and uniqueness results are provided in SM1.2.

\begin{remark}[Interpretation of the uniqueness conditions] Condition \eqref{eq:epe1} requires the saturation threshold for T cell activation to be no smaller than those associated with macrophage recruitment and polarization. Condition \eqref{eq:epe2} requires the macrophage-mediated increase in tumor proliferation to be sufficiently weak relative to the density-dependent limitation imposed by the tumor carrying capacity. These conditions are sufficient for uniqueness and describe a regime in which tumor-immune feedback does not generate multiple positive equilibria. 
\end{remark}

The existence of a positive equilibrium $E^* = (C^*, T^*, A^*)$ indicates that the tumor may persist in the host. We next examine whether this coexistence state is locally stable. 

\begin{theorem}[Local asymptotic stability of the positive equilibrium]
	\label{thm:PE_stability}
	Assume that $\alpha > d_C$ and that conditions \eqref{eq:epe1} and \eqref{eq:epe2} hold. If
	\begin{equation}
		\label{eq:condition_stability}
		\frac{ \omega d_A}{\omega + \beta } > \frac{ \delta \beta }{d_T}  \left(\frac{ \chi }{ \sigma }  +  \frac{\gamma M }{ \theta  }  \right),
	\end{equation}
	then the unique positive equilibrium $E^*$ is locally asymptotically stable.
\end{theorem}

\begin{proof} 
	For convenience, let
	\begin{equation*}
		\begin{aligned}
			x & = \frac{\alpha}{\beta}(1 + \epsilon A^*) C^*, & y &= \eta C^*, & z &= \alpha\epsilon\left(1-\frac{C^*}{\beta}\right)C^*,\\
			u&=\frac{\kappa N\omega}{(\omega+C^*)^2},&
			v&=\delta A^*+d_T,&
			w&=\delta T^*,\\
			h&=\frac{\chi\sigma}{(\sigma+C^*)^2}
			+\frac{\gamma M\theta}{(\theta+C^*)^2}.
		\end{aligned}
	\end{equation*}
	The Jacobian matrix at $E^*$ is 
	\begin{equation*}
		J(E^*)=
		\begin{pmatrix}
			-x & -y & z\\
			u & -v & -w\\
			h & 0 & -d_A
		\end{pmatrix}.
	\end{equation*}
	Its characteristic polynomial is
	\begin{equation}
		\label{eq:re_ap12}
		\lambda^3 + a_1 \lambda^2 + a_2 \lambda +a_3=0,
	\end{equation}
	where
	\begin{equation*}
		\begin{aligned}
			a_1&=x+v+d_A,\\
			a_2&=x(v+d_A)+vd_A+yu-zh,\\
			a_3&=xvd_A+yud_A-ywh-zvh.
		\end{aligned}
	\end{equation*}
	Clearly, $a_1>0$.
	
	Condition \eqref{eq:epe2} implies
	\begin{equation}
		\label{eq:re_ap_a}
		\frac{d_A}{\beta}
		-\epsilon\left(\frac{\chi}{\sigma}+\frac{\gamma M}{\theta}\right)
		\geq0.
	\end{equation}
	Since $1+\epsilon A^*\geq1$,
	$1-C^*/\beta<1$, and $h\leq\frac{\chi}{\sigma}+\frac{\gamma M}{\theta}$, we obtain
	\begin{equation}
		\begin{aligned}
			a_3 \geq {} \alpha C^*v \left[
			\frac{d_A}{\beta} -\epsilon\left(\frac{\chi}{\sigma}+\frac{\gamma M}{\theta}\right)
			\right] +\eta C^*\left[
			\frac{\kappa N\omega d_A}{(\omega+C^*)^2} - \delta T^*
			\left(\frac{\chi}{\sigma}+\frac{\gamma M}{\theta}\right)
			\right].
		\end{aligned}
	\end{equation}
	At equilibrium,
	\begin{equation}
		T^*=\frac{\kappa NC^*}{(\omega+C^*)(\delta A^*+d_T)} < \frac{\kappa NC^*}{(\omega+C^*)d_T}.
	\end{equation}
	Therefore,
	\begin{equation}
		\begin{aligned}
			\frac{\kappa N\omega d_A}{(\omega+C^*)^2} - \delta T^* \left(\frac{\chi}{\sigma}+\frac{\gamma M}{\theta}\right) >
			\frac{\kappa N}{\omega+C^*} \left[ \frac{\omega d_A}{\omega+C^*} -\frac{\delta C^*}{d_T} \left(\frac{\chi}{\sigma}+\frac{\gamma M}{\theta}\right) \right].
		\end{aligned}
	\end{equation}
	Since $0<C^*<\beta$ from the proof of Theorem \ref{thm:PE_existence}, condition \eqref{eq:condition_stability} implies that the right-hand side is positive. Hence, $a_3>0$.
	
	Finally, a direct calculation yields
	\begin{equation*}
		\begin{aligned}
			a_1a_2-a_3
			={}&x(v+d_A)(x+v+d_A)+vd_A(v+d_A)+yu(x+v)\\
			&+h\bigl[yw-z(x+d_A)\bigr].
		\end{aligned}
	\end{equation*}
	Dropping the positive terms $vd_A(v+d_A)$, $yu(x+v)$, and $hyw$ gives
	\begin{equation*}
		\begin{aligned}
			a_1a_2-a_3
			>{}&x(v+d_A)(x+v+d_A)-zh(x+d_A)\\
			>{}&\alpha C^*(x+d_A)
			\left[
			\frac{d_A}{\beta}
			-\epsilon\left(\frac{\chi}{\sigma}+\frac{\gamma M}{\theta}\right)
			\right]\geq0,
		\end{aligned}
	\end{equation*}
	where the last inequality follows from \eqref{eq:re_ap_a}. Thus,
	$a_1a_2-a_3>0$.
	
	The Routh-Hurwitz criterion now implies that all eigenvalues of $J(E^*)$ have negative real parts. Hence, $E^*$ is locally asymptotically stable.
\end{proof}

Numerical verification of Theorem \ref{thm:PE_stability} is provided in SM1.3.

\begin{remark}[Hopf bifurcation] 
Let $\mathcal{H} = a_1 a_2 - a_3$. For the positive equilibrium, the Routh-Hurwitz conditions are $a_1>0$, $a_3>0$, and $\mathcal{H}>0$. Suppose that a model parameter $\xi$ has a critical value $\xi_0$ such that $a_1(\xi_0) > 0$, $a_2(\xi_0)>0$, $a_3(\xi_0) > 0$, $\mathcal{H}(\xi_0) = 0$. Then the Jacobian matrix at $E^*$ has a pair of purely imaginary eigenvalues and one negative real eigenvalue. If, in addition, $\left. \frac{\mathrm{d} \mathcal{H}}{\mathrm{d} \xi} \right|_{\xi = \xi_0}\neq 0$, the conjugate pair crosses the imaginary axis transversally. Thus, $\xi_0$ is a candidate Hopf bifurcation point. The existence and direction of the corresponding periodic solutions can be determined by numerical continuation or by computing the first Lyapunov coefficient.
\end{remark}

In this work, we use the activation rate $\kappa$ as the bifurcation parameter, since it directly controls the strength of the antitumor immune response. The numerical bifurcation analysis is presented in SM1.4.

\section{Results}

\subsection{Model calibration and validation of tumor growth dynamics}

We calibrated the model using tumor growth data from three mouse cohorts: immunodeficient mice, immunocompetent mice, and arginine-treated immunocompetent mice \cite{Zhu.CancerCell.2025}. To relate the experimental measurements to the model variable $C$, tumor volume $V$ was calculated from the tumor cell number according to 
\begin{equation}
V = \frac{V_{\rm{cell}} \times C}{V_{\rm{fraction}}},
\end{equation}
where $V_{\rm{cell}} = 2.572 \times 10^{-6}\ \mathrm{mm}^3/\mathrm{cell}$ is the volume of a single tumor cell and $V_{\rm{fraction}}=0.37$ is the cellular volume fraction \cite{Wang.NPJPrecisOncol.2023,Wang.JImmunotherCancer.2021}.

The model contains 16 parameters. Based on previous studies, 10 parameters were fixed, and the remaining six parameters were estimated from the experimental data (see SM2.1). Specifically, the unknown parameter vector $\boldsymbol{p}$ was determined by maximizing the coefficient of determination ($R^2$) between model predictions and observations: 
\begin{equation}
\max_{\boldsymbol{p}} J(\boldsymbol{p}) = 1-\frac{\sum_{i=1}^{n} \left[y(t_i)-\hat{y}(t_i;\boldsymbol{p})\right]^2} {\sum_{i=1}^{n}\left[y(t_i)-\bar{y}\right]^2},
\end{equation}
where $\hat{y}(t_i; \boldsymbol{p})$ and $y(t_i)$ denote the model prediction and experimental observation at time $t_i$, respectively, and $\bar{y}$ is the mean of the $n$ observations.

\begin{figure}[htbp]
	\centering
	\includegraphics[width=12.5cm]{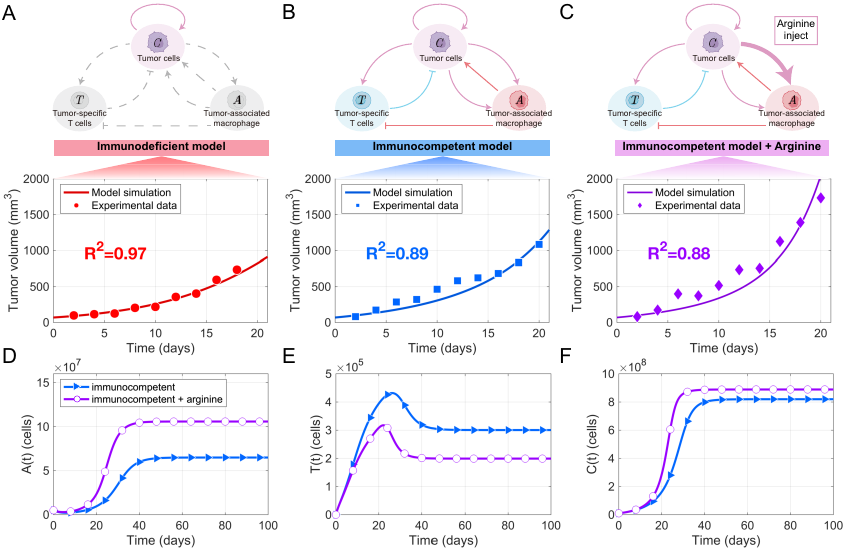}
	\caption{Model calibration against experimental tumor growth data. \textbf{A}-\textbf{C} Schematic mechanisms and tumor growth curves for the immunodeficient, immunocompetent, and arginine-treated cohorts, respectively. Experimental data are from \cite{Zhu.CancerCell.2025}. \textbf{D}-\textbf{F} Temporal dynamics of tumor-associated macrophages, tumor-specific T cells, and tumor cells, respectively. }
	\label{Fig2}
\end{figure}

The model was configured for each cohort as follows. For the immunodeficient cohort, immune components were assumed to be absent, so that $T(t)=A(t)=0$. The immunocompetent cohort was simulated using the baseline model. For the arginine-treated cohort, the macrophage polarization rate was set to $\gamma = 0.5\ \mathrm{day}^{-1}$ to represent the arginine-induced phenotypic transition. The calibrated model agreed well with the experimental data, with $R^2$ values of $0.97$, $0.89$, and $0.88$ for the immunodeficient, immunocompetent, and arginine-treated cohorts, respectively (Fig.~\ref{Fig2}A-C). 

The simulations indicate more rapid tumor growth in immunocompetent mice than in immunodeficient mice, and arginine injection further increases tumor growth. This pattern results from two coupled mechanisms. On the one hand, tumor cells recruit circulating macrophages through chemokine-mediated paracrine signaling. On the other hand, arginine facilitates the phenotypic polarization of macrophages. The resulting tumor-associated macrophages support tumor progression through the secretion of tumor-promoting factors. Although tumor-specific T cells are activated in response to tumor cells, their antitumor activity is attenuated by the immunosuppressive environment generated by tumor-associated macrophages.

Consistent with these mechanisms, arginine induces a rapid increase in tumor-associated macrophages, which subsequently approach a higher steady-state level than in the untreated immunocompetent cohort (Fig.~\ref{Fig2}D). In contrast, tumor-specific T cells in the arginine-treated cohort reach a lower peak and settle at a lower steady-state level than those in the untreated cohort (Fig.~\ref{Fig2}E). Consequently, the model predicts weaker immune control of the tumor, leading to faster tumor growth and a larger final tumor burden (Fig.~\ref{Fig2}F). A global sensitivity analysis based on the Sobol' method is provided in SM3.

\subsection{Deciphering regulatory mechanisms of tumor-immune interactions}

\begin{figure}[htbp]
	\centering
	\includegraphics[width=12.5cm]{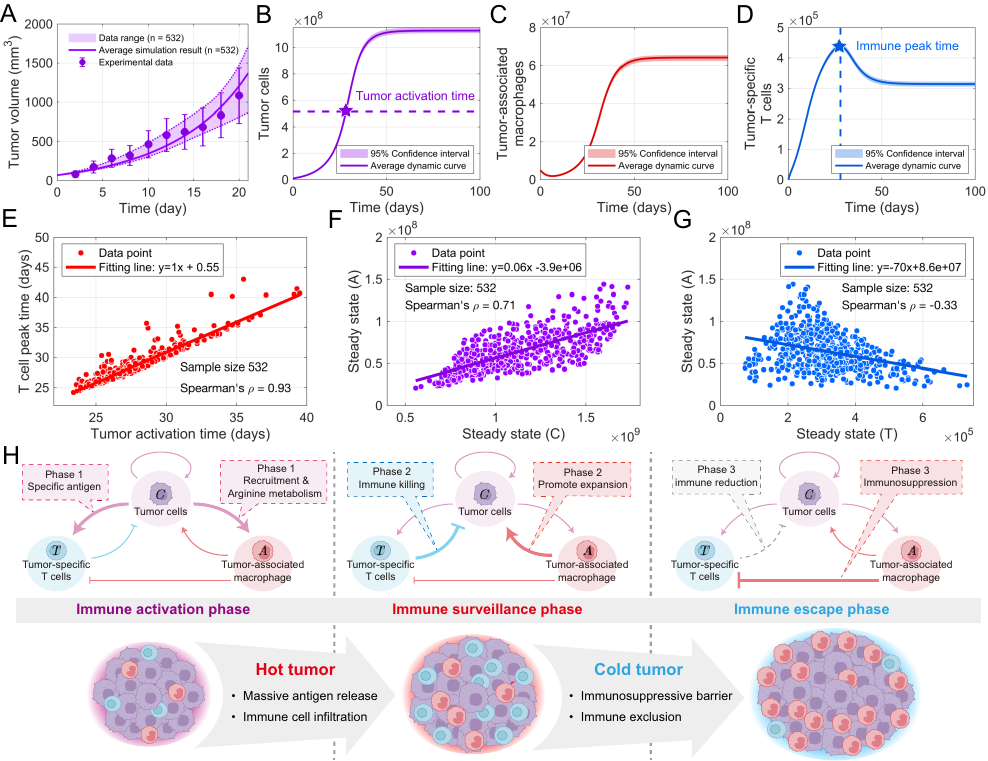}
	\caption{Analysis of tumor-immune interactions and population dynamics using the TVInfABC method. \textbf{A} Experimental tumor growth data \cite{Zhu.CancerCell.2025} and the range of simulated tumor trajectories obtained from the accepted virtual samples. \textbf{B}-\textbf{D} Temporal dynamics of tumor cells (\textbf{B}), tumor-associated macrophages (\textbf{C}), and tumor-specific T cells (\textbf{D}). \textbf{E}-\textbf{G} Correlation analyses between tumor activation time and T cell peak time (\textbf{E}), tumor and tumor-associated macrophage steady-state values (\textbf{F}), and tumor-specific T cell and tumor-associated macrophage steady-state levels (\textbf{G}). \textbf{H} Schematic illustration of the proposed regulatory mechanism of tumor-immune interactions.}
	\label{Fig4}
\end{figure}

To characterize the regulatory mechanisms underlying tumor-immune interactions, we used the Tumor Volume Inference using Approximate Bayesian Computation (TVInfABC) method to generate a virtual cohort \cite{Li.DCDSI.2026,Li.BullMathBiol.2024,Li.NPJSystBiolAppl.2025} (see SM4). We sampled heterogeneous parameter combinations representing key biological processes, including tumor-intrinsic properties (the proliferation rate $\alpha$ and the carrying capacity $\beta$), macrophage-related processes (the arginine-mediated polarization rate $\gamma$ and baseline macrophage level $M$), and T cell-mediated immunity (the killing rate $\eta$ and baseline T cell level $N$). Among 10,000 sampled parameter combinations, 532 produced tumor growth trajectories with good agreement with the experimental data (Fig.~\ref{Fig4}A; see SM4). Across these accepted virtual samples, tumor-associated macrophages gradually accumulated as the tumor expanded, whereas tumor-specific T cells initially increased and subsequently declined (Fig.~\ref{Fig4}B-D).

To quantify the temporal relationship between tumor progression and immune activation, we introduced two dynamic indices: tumor activation time and T cell peak time. The tumor activation time was defined as the first time at which the tumor cell population reached 40\% of its steady-state value (Fig.~\ref{Fig4}B):
\begin{equation}
\Phi := \inf\left\{ t \geq 0: C(t) \geq 0.4 \times C^*\right\},
\end{equation}
where $C^*$ denotes the tumor steady-state value. The T cell peak time was defined as the earliest time at which tumor-specific T cells attained their maximum value over the simulation interval (Fig.~\ref{Fig4}D):
\begin{equation}
\Psi := \inf \left\{ t \in [0, t_{\mathrm{end}}]: T(t) = \max_{\tau\in[0, t_{\mathrm{end}}]} T(\tau)\right\} .
\end{equation}
A strong positive correlation was observed between tumor activation time and T cell peak time, with Spearman's $\rho = 0.93$ (Fig.~\ref{Fig4}E). In addition, tumor and tumor-associated macrophage steady-state levels were positively correlated ($\rho = 0.71$; Fig.~\ref{Fig4}F), whereas tumor-specific T cell and tumor-associated macrophage steady-state levels were negatively correlated ($\rho = -0.33$; Fig.~\ref{Fig4}G).

Together, these results suggest a three-phase progression of tumor-immune interactions (Fig.~\ref{Fig4}H). During the early immune activation phase, tumor expansion increases tumor antigen availability, promoting the activation of tumor-specific T cells, while arginine-related processes promote macrophage polarization. During the intermediate immune surveillance phase, the increased T cell response restrains tumor growth, corresponding to a relatively immune-active (``hot'') tumor state. During the late immune escape phase, tumor-associated macrophages accumulate in the tumor microenvironment and promote immunosuppression, thereby attenuating T cell activity and favoring an immune-suppressed (``cold'') tumor state. Thus, the model describes a transition from immune activation and surveillance to immune escape. In particular, macrophage recruitment and polarization emerge as key processes associated with the disruption of the tumor-immune balance and the transition from a hot to a cold tumor state.

\subsection{Quantitative evaluation and stratification of combination therapies}

\begin{figure}[htbp]
	\centering
	\includegraphics[width=12.5cm]{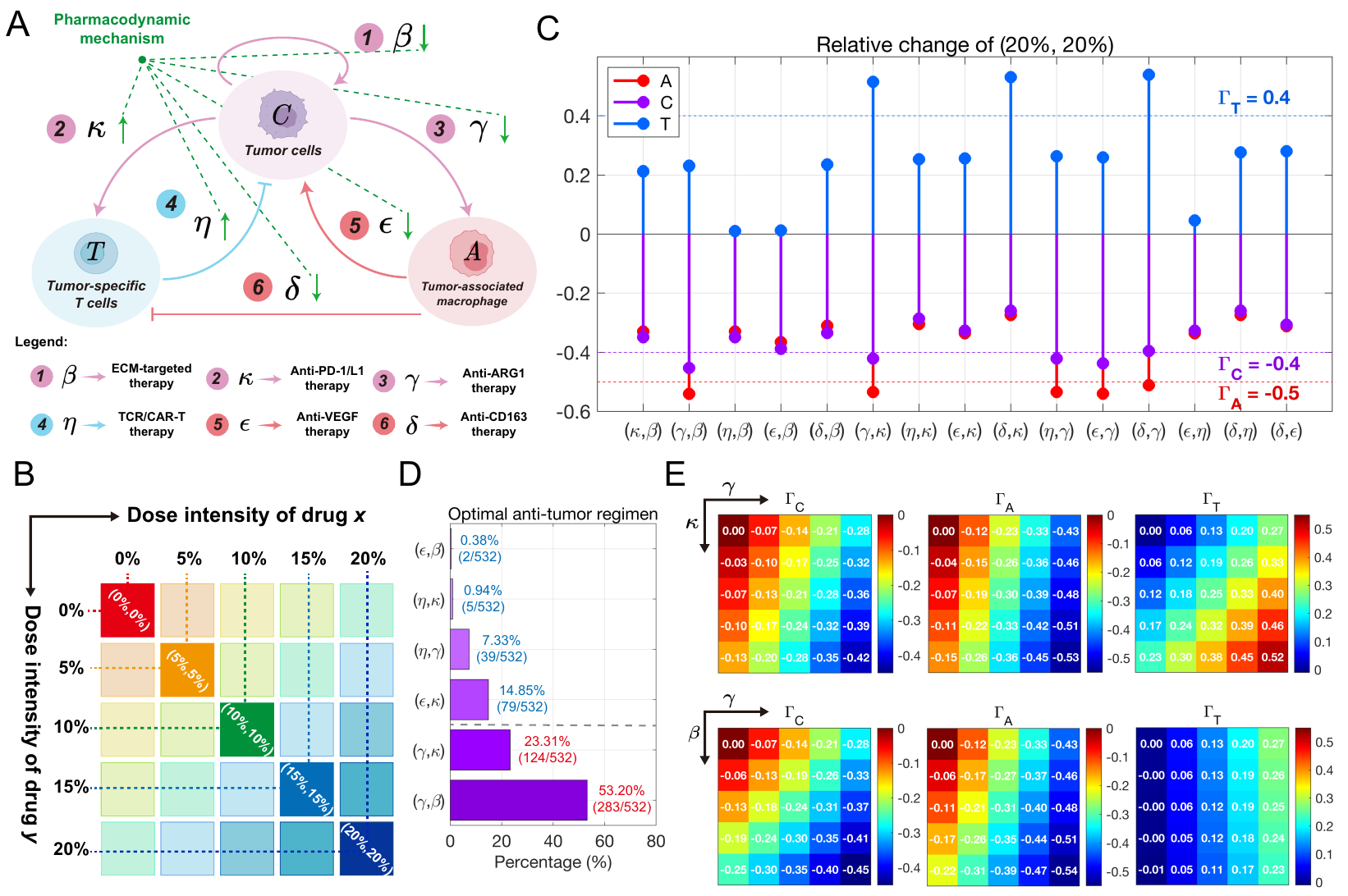}
	\caption{Quantitative evaluation of combination therapies in the virtual cohort. \textbf{A} Tumor-immune regulatory network and parameter-based representations of candidate therapeutic interventions. Decreases in $\beta$, $\gamma$, $\epsilon$, and $\delta$ represent ECM-targeted, anti-ARG1, anti-VEGF, and anti-CD163 therapies, respectively, whereas increases in $\kappa$ and $\eta$ represent anti-PD-1/PD-L1 and TCR/CAR-T therapies. \textbf{B} Schematic illustration of combination treatment simulations based on simultaneous perturbation of two parameters. \textbf{C} Mean relative changes in tumor-associated macrophages, tumor cells, and tumor-specific T cells across the 532 virtual samples under a dose combination of $(20\%, 20\%)$. \textbf{D} Distribution of optimal treatment strategies across the 532 virtual samples. \textbf{E} Dose-dependent relative changes under the $(\gamma, \kappa)$ and $(\gamma, \beta)$ combinations.}
	\label{Fig5}
\end{figure}

To evaluate the effect of combination therapies across heterogeneous tumor-immune conditions, we performed a systematic perturbation analysis of drug-relevant pathways in the tumor-immune regulatory network (Fig.~\ref{Fig5}A; see SM5). The virtual cohort was represented by the 532 parameter combinations accepted by the TVInfABC procedure. For each virtual sample, the parameters $\alpha$, $\beta$, $\gamma$, $\eta$, $N$, and $M$ were taken from the corresponding accepted parameter set. To account for uncertainty in the remaining drug-relevant parameters, $\kappa$, $\epsilon$, and $\delta$ were independently perturbed within 5\% of their baseline values.

Drug interventions were represented by parameter changes relative to the untreated baseline. For an intervention targeting parameter $p_i$, the corresponding dose intensity is denoted by $D_i \in [0, 1]$ and represents the relative magnitude of the treatment-induced change from the untreated baseline value $p_{i,0}$. For interventions that downregulate a parameter, the treated value was defined as $p_i(D_i) = p_{i,0}(1 - D_i)$, whereas for interventions that upregulate a parameter, it was defined as $p_i(D_i) = p_{i,0} (1 + D_i)$. Accordingly, decreases in $\beta$, $\gamma$, $\epsilon$, and $\delta$ represented ECM-targeted, anti-ARG1, anti-VEGF, and anti-CD163 interventions, respectively, whereas increases in $\kappa$ and $\eta$ represented anti-PD-1/PD-L1 and TCR/CAR-T therapies, respectively (Fig.~\ref{Fig5}A). Here, downregulation of the parameter $\gamma$ represents inhibition of arginine-mediated macrophage polarization. The biological rationale for these parameter-treatment mappings is provided in SM5.

For a treatment combination with dose intensities $D_i$ and $D_j$, we defined the relative change in the population of cell type $x \in \{C, T, A\}$ at the simulation endpoint $t_{\mathrm{end}} = 35$ days as follows  (Fig.~\ref{Fig5}B) 
\begin{equation}
\label{eq:Gamma}
\Gamma_x(D_i, D_j) = \frac{Q_{x}(D_i, D_j) - Q_{x}(0, 0)}{Q_{x}(0, 0)},\ x \in \{C, T, A\},
\end{equation}
where $Q_x(D_i, D_j)$ denotes the predicted abundance of cell type $x$ at day 35 under treatment $(D_i, D_j)$, and $Q_x(0, 0)$ is the corresponding untreated value. Positive and negative values of $\Gamma_x$ indicate an increase and a decrease, respectively, relative to the untreated baseline.

At the population level, the treatment combinations of $(\gamma, \beta)$, $(\gamma, \kappa)$, $(\eta, \gamma)$, $(\epsilon, \gamma)$, and $(\delta, \gamma)$ produced lower mean tumor burdens and tumor-associated macrophage levels than the untreated baseline (Fig.~\ref{Fig5}C; see SM5). These results consistently identify inhibition of arginine-mediated macrophage polarization as an important component of effective combination therapies. In particular, combining the reduction of $\gamma$ with interventions targeting tumor growth, T cell activation, T cell killing, macrophage-mediated immunosuppression, or arginine-enhanced tumor proliferation improved tumor control across the virtual cohort. This result is consistent with evidence that arginase-1 inhibition can improve the efficacy of cancer immunotherapy \cite{Cane.SciTranslMed.2023}. 

The treatment combinations of $(\gamma, \kappa)$, $(\delta, \kappa)$, and $(\delta, \gamma)$ also increased the mean tumor-specific T cell level (Fig.~\ref{Fig5}C). This pattern suggests that enhancing T cell activation or relieving macrophage-mediated immunosuppression may improve antitumor immunity, particularly when combined with inhibition of macrophage polarization.

For each virtual sample, we defined the optimal strategy as the combination that minimized the tumor population at day 35 among the candidate treatment regimens. Under this criterion, $53.20\%$ and $23.31\%$ of virtual samples achieved their best tumor control under the $(\gamma, \beta)$ and $(\gamma, \kappa)$ treatment combinations, respectively (Fig.~\ref{Fig5}D). Although the $(\gamma, \beta)$ combination was more frequently identified as the optimal strategy, the $(\gamma, \kappa)$ combination produced a greater increase in tumor-specific T cells over a range of dose intensities (Fig.~\ref{Fig5}E). Consistently, experimental studies confirm that combination regimens of anti-ARG1 with either anti-PD-1 or ECM-targeted therapy provide superior control of tumor progression \cite{Aaboe-Jorgensen.CancerImmunolRes.2021,Lauwers.PNAS.2024,Puttock.NatCommun.2023}, which provides qualitative support for the model-based predictions. Collectively, these findings provide a theoretical rationale for investigating precision combination therapies grounded in dual metabolic-immune regulation and identify candidate functional nodes linking metabolic reprogramming to immune escape within the tumor microenvironment.

\begin{figure}[htbp]
\centering
\includegraphics[width=12.5cm]{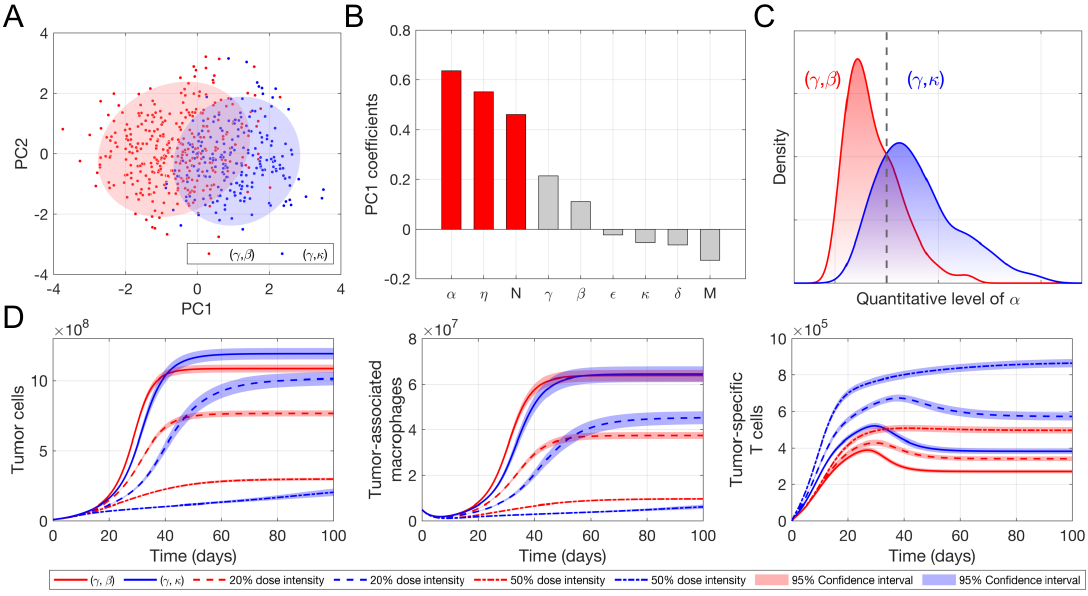}
\caption{Parameter features associated with optimal treatment strategies. \textbf{A} Principal component analysis of virtual samples classified according to their optimal treatment strategy. Red and blue points denote samples for which $(\gamma, \beta)$ and $(\gamma, \kappa)$ treatment combinations, respectively, achieved the greatest tumor reduction. \textbf{B} Contributions of model parameters to the first principal component. \textbf{C} Probability-density distributions of the tumor proliferation rate $\alpha$ in samples for which the $(\gamma,\beta)$ or $(\gamma,\kappa)$ combination was identified as optimal. \textbf{D} Representative tumor-immune dynamics under different treatment intensities for the corresponding virtual sample groups.}
\label{Fig6}
\end{figure}

To identify the key factors associated with these distinct optimal strategies, we performed principal component analysis of the virtual samples classified according to their best treatment outcomes (Fig.~\ref{Fig6}A). The tumor proliferation rate $\alpha$ was the largest contributor to the first principal component and was the major source of variation between treatment-response groups (Fig.~\ref{Fig6}B). Virtual samples with higher values of $\alpha$ were more likely to benefit from the $(\gamma, \kappa)$ combination, whereas those with lower values of $\alpha$ were more likely to benefit from the $(\gamma, \beta)$ combination (Fig.~\ref{Fig6}C). In addition, a higher immune killing rate $\eta$ and higher baseline T cell levels $N$ were associated with improved responses to the $(\gamma, \kappa)$ combination strategy (Fig.~\ref{Fig6}B; see SM6).

Representative simulations further showed that samples with stronger T cell activation exhibited greater early sensitivity to the $(\gamma, \kappa)$ combination (Fig.~\ref{Fig6}D). Increasing treatment intensity reduced tumor-associated macrophage levels and promoted the expansion of tumor-specific T cells in these samples. Together, these analyses suggest that tumor proliferative capacity and baseline immune competence may help stratify the virtual cohort and guide the selection between therapies targeting tumor growth and those enhancing T-cell activation.

\subsection{Quantification of dose-dependent combination effects}

To quantify dose-dependent treatment efficacy and evaluate potential interactions between therapeutic interventions, we fitted single-agent and combination dose-response relationships to the numerical simulation results. We then used the Bliss independence model to assess whether the observed combination efficacy differed from that expected from the corresponding monotherapies. 

\subsubsection{Single-agent dose-response relationships}

For each monotherapy and each relative dose intensity $D_i$, we first simulated the model and quantified the corresponding treatment efficacy by the relative reduction in tumor burden:
\begin{equation}
\label{eq:4.7}
\varepsilon_i^{\mathrm{sim}}(D_i) = -\Gamma_C(D_i),\ i \in \{\gamma, \kappa, \beta\}
\end{equation}
where $\Gamma_C(D_i)$ is the relative change in the tumor cell population, defined analogously to Eq.~\eqref{eq:Gamma}. Thus, larger values of $\varepsilon_i^{\mathrm{sim}}(D_i)$ indicate greater antitumor efficacy.

We then fitted the simulated dose-response data using the Hill-type equation \cite{Sen.TrendsPharmacolSci.2019,Meyer.CellSyst.2019}:
\begin{equation}
\label{eq:One}
\varepsilon_i(D_i) = \frac{\varepsilon_i^{\max} \left(D_i/D_{50,i}\right)^{n_i}}{1+\left(D_i/D_{50,i}\right)^{n_i}},
\end{equation}
where $\varepsilon_i^{\max}$ is the fitted maximal efficacy, $D_{50, i}$ is the dose intensity required to achieve one-half of $\varepsilon_i^{\max}$, and $n_i$ is the Hill coefficient. The parameter set $(\varepsilon_i^{\max}, D_{50, i}, n_i)$ was estimated by fitting Eq. ~\eqref{eq:One} to $\varepsilon_i^{\mathrm{sim}}(D_i)$; see SM2.2 for details.

\begin{figure}[h!]
	\centering
	\includegraphics[width=12.5cm]{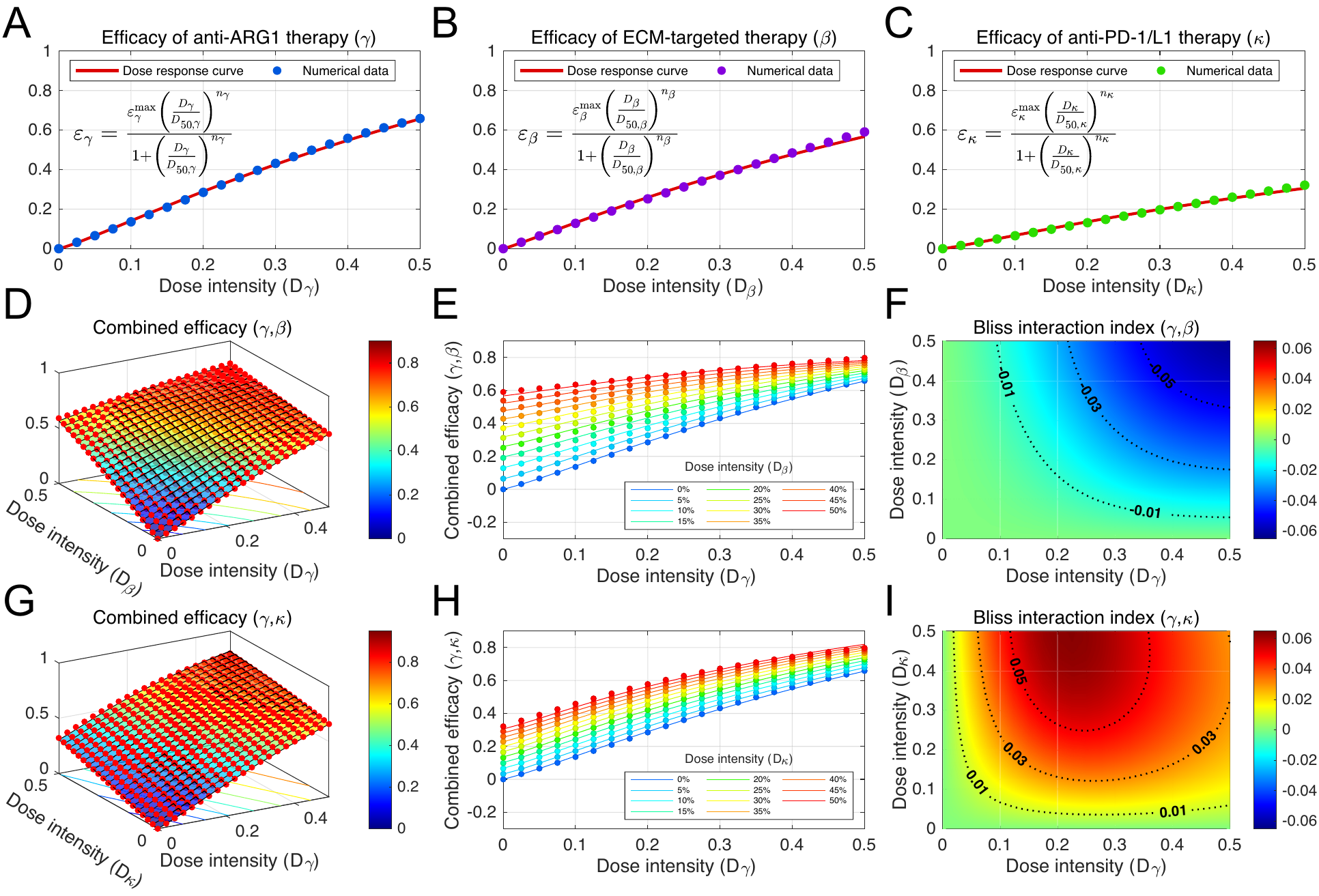}
	\caption{Dose-dependent efficacy and Bliss interaction analysis of combination therapies. \textbf{A}-\textbf{C} Simulated and fitted single-agent dose-response relationships for anti-ARG1 therapy, ECM-targeted therapy, and anti-PD-1/PD-L1 therapy, corresponding to interventions targeting $\gamma$, $\beta$, and $\kappa$, respectively. \textbf{D} and \textbf{G} Fitted combined-efficacy surfaces for the $(\gamma, \beta)$ and $(\gamma, \kappa)$ treatment combinations, respectively. \textbf{E} and \textbf{H} Combined efficacy as a function of anti-ARG1 dose intensity $D_\gamma$ at different dose intensities of ECM-targeted therapy ($D_\beta$) and anti-PD-1/PD-L1 therapy ($D_\kappa$), respectively. \textbf{F} and \textbf{I} Bliss interaction indices for the $(\gamma, \beta)$ and $(\gamma, \kappa)$ treatment combinations, respectively. Positive values indicate synergy, negative values indicate antagonism, and zero indicates agreement with the Bliss independence prediction. Both panels use the same color scale.}
	\label{Fig7}
\end{figure}

At a dose intensity of $D_\gamma = 0.5$, anti-ARG1 therapy reduced tumor burden by more than $60\%$ (Fig.~\ref{Fig7}A). At dose intensities of $D_\beta = 0.5$ and $D_\kappa = 0.5$, ECM-targeted therapy and anti-PD-1/PD-L1 therapy reduced tumor burden by approximately $60\%$ and $40\%$, respectively (Fig.~\ref{Fig7}B-C). These results indicate that anti-ARG1 therapy has the strongest single-agent effect among the three interventions over the considered dose range.

\subsubsection{Combined dose-response relationships}

For each two-drug combination $(i, j)$ and each pair of relative dose intensities ($D_i, D_j)$, we simulated the model and quantified the combination efficacy by the relative reduction in tumor burden:
\begin{equation}
\varepsilon_{ij}^{\mathrm{sim}}(D_i,D_j) = -\Gamma_C(D_i,D_j),
\end{equation}
where $\Gamma_C(D_i, D_j)$ is defined as in Eq.~\eqref{eq:Gamma}. Larger values of $\varepsilon_{ij}^{\mathrm{sim}}(D_i, D_j)$ indicate greater antitumor efficacy of the combination.

We then fitted the simulated combination dose-response data using an extension of the single-agent Hill-type model that accounts for treatment-induced changes in drug potency and interactions in the joint response surface \cite{Sen.TrendsPharmacolSci.2019,Meyer.CellSyst.2019}:
\begin{equation}
\label{eq:Two}
\varepsilon_{ij}(D_i,D_j) = \frac{\varepsilon_i^{\max}x_i+\varepsilon_j^{\max}x_j + \bigl(\varepsilon_i^{\max}+\varepsilon_j^{\max} -\varepsilon_i^{\max}\varepsilon_j^{\max}\bigr)(1+\zeta_{ij})x_i x_j}{1+x_i+x_j+(1+\zeta_{ij})x_i x_j},
\end{equation}
where 
$$x_i = \left(\frac{D_i}{D_{50,ij}}\right)^{n_i}, \quad x_j = \left(\frac{D_j}{D_{50,ji}}\right)^{n_j}, $$
and
$$D_{50,ij} = \frac{D_{50,i}}{1+\xi_{ij}\frac{D_j}{D_{50,j}}}, \quad D_{50,ji} = \frac{D_{50,j}}{1+\xi_{ji}\frac{D_i}{D_{50,i}}}.$$
Here, the single-agent parameters $(\varepsilon_i^{\max}, D_{50, i}, n_i)$ and $(\varepsilon_j^{\max}, D_{50, j}, n_j)$ were obtained from the corresponding monotherapy dose-response fits. The parameters $\xi_{ij}$ and $\xi_{ji}$ quantify changes in the effective potency of treatments $i$ and $j$ induced by their partner treatment, respectively, whereas $\zeta_{ij}$ characterizes the interaction in the joint response surface. These parameters are distinct from those of the tumor-immune ODE model.

The parameter set $(\xi_{ij}, \xi_{ji}, \zeta_{ij})$ was estimated by fitting Eq.~\eqref{eq:Two} to $\varepsilon_{ij}^{\mathrm{sim}}(D_i, D_j)$; see SM2.2 for details. The fitted response surfaces closely reproduced the simulation results over the examined dose range (Fig.~\ref{Fig7}D, E, G, and H). At the upper end of this range, both the $(\gamma, \beta)$ and $(\gamma, \kappa)$ combinations achieved approximately $90\%$ tumor reduction (Fig.~\ref{Fig7}D and G). 

Cross-sectional views of the fitted response surfaces showed that the combined efficacy of both $(\gamma, \beta)$ and $(\gamma, \kappa)$ treatments increased as the dose intensity of either intervention increased (Fig.~\ref{Fig7}E and H). For the $(\gamma, \beta)$ combination, the additional benefit of increasing anti-ARG1 dose intensity $D_\gamma$ became less pronounced at higher ECM-targeted therapy dose intensities $D_\beta$, and the response curve converged at high $D_\gamma$ values (Fig.~\ref{Fig7}E), suggesting saturation of the tumor-reduction effect. In contrast, the $(\gamma, \kappa)$ combination maintained a more pronounced increase in efficacy over the examined dose range (Fig.~\ref{Fig7}H), with the largest treatment effects observed at higher values of both $D_{\gamma}$ and $D_{\kappa}$. These distinct dose-response patterns were subsequently evaluated relative to the Bliss independence reference model.

\subsubsection{Bliss interaction analysis}

According to the Bliss independence model, which assumes that two interventions act independently without pharmacodynamic interaction, the expected combined efficacy is 
\begin{equation}
\label{eq:bliss-prediction}
\varepsilon_{\mathrm{Predict}} = \varepsilon_i+\varepsilon_j-\varepsilon_i\varepsilon_j.
\end{equation}
Here, $\varepsilon_i$ and $\varepsilon_j$ denote the efficacies of the individual interventions. We quantified the interaction between the two interventions using the Bliss interaction index, defined as the difference between the observed combination efficacy, $\varepsilon_{\mathrm{Observed}} = \varepsilon_{ij}$, and the efficacy predicted under Bliss independence \cite{Sen.TrendsPharmacolSci.2019,Meyer.CellSyst.2019}:
\begin{equation}
\label{eq:bliss-index}
\begin{aligned}
\beta_{\mathrm{Bliss}} = \varepsilon_{\mathrm{Observed}} - \varepsilon_{\mathrm{Predict}} = \varepsilon_{ij} - \left(\varepsilon_i+\varepsilon_j -\varepsilon_i\varepsilon_j\right).
\end{aligned}
\end{equation}
A positive value of $\beta_{\mathrm{Bliss}}$ indicates synergy, whereas a negative value indicates antagonism. A value of $\beta_{\mathrm{Bliss}} = 0$ indicates that the observed combination efficacy is consistent with the Bliss independence prediction.

The $(\gamma, \beta)$ combination exhibited negative Bliss interaction indices over the nonzero combination dose range, indicating antagonism relative to Bliss independence (Fig.~\ref{Fig7}F). The Bliss interaction index became more negative as both $D_\gamma$ and $D_\beta$ increased. In contrast, the $(\gamma, \kappa)$ combination exhibited positive Bliss interaction indices over the nonzero combination dose range, indicating synergy (Fig.~\ref{Fig7}I). The strongest synergy was observed at intermediate anti-ARG1 dose intensities ($D_\gamma \approx 0.2$-$0.3$) and relatively high values of the anti-PD-1/PD-L1 dose intensity $D_\kappa$.

\subsection{Bifurcation analysis}

To investigate how the tumor-specific T cell activation rate $\kappa$ and the tumor-associated macrophage polarization rate $\gamma$ affect the long-term system dynamics, we numerically computed the positive steady states of the model and determined their stability. In particular, we examined whether variations in these two parameters can generate bistability.

\begin{figure}[htbp]
	\centering
	\includegraphics[width=12.5cm]{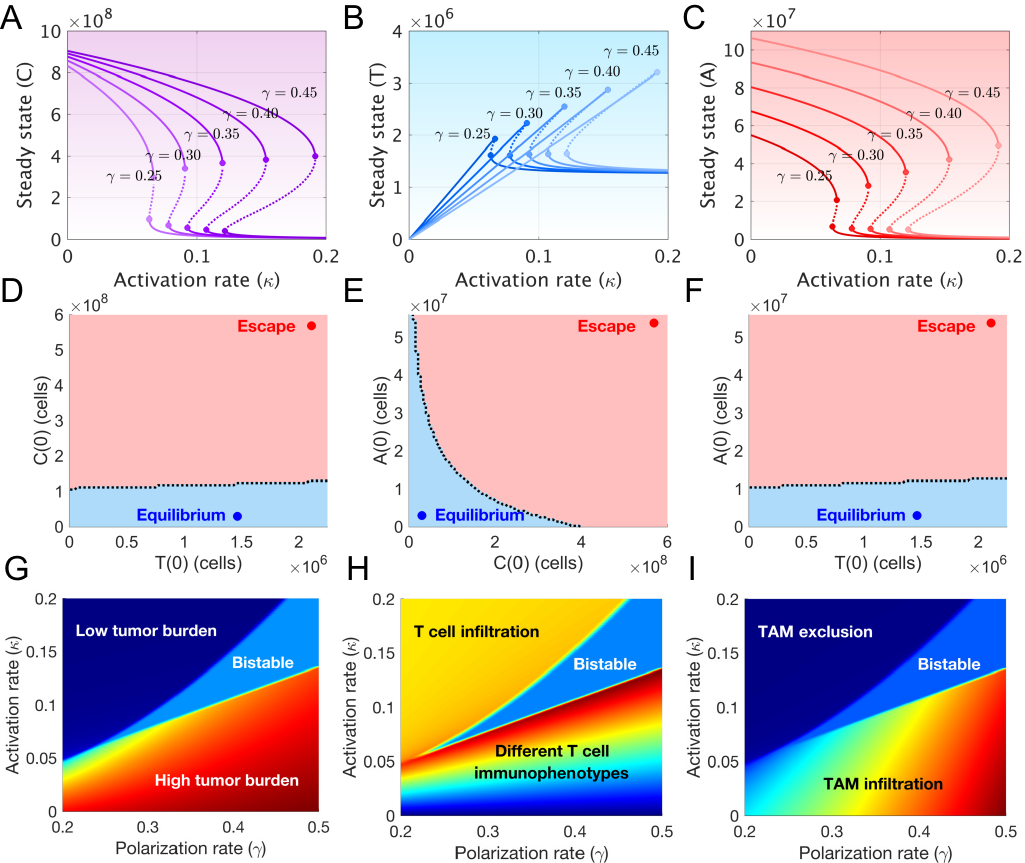}
	\caption{Bifurcation analysis with respect to the tumor-specific T cell activation rate $\kappa$ and the tumor-associated macrophage polarization rate $\gamma$. \textbf{A}-\textbf{C} One-parameter bifurcation diagrams showing the steady-state populations of tumor cells (\textbf{A}), tumor-specific T cells (\textbf{B}), and tumor-associated macrophages (\textbf{C}) as functions of $\kappa$. The polarization rate $\gamma$ was fixed at $0.25$, $0.30$, $0.35$, $0.40$, and $0.45$, respectively. Solid and dashed curves indicate stable and unstable equilibria, respectively. \textbf{D}-\textbf{F} Basins of attraction under different initial conditions for $\kappa = 0.1$ and $\gamma = 0.35$. \textbf{G}-\textbf{I} Two-parameter bifurcation diagrams showing the steady-state populations of tumor cells (\textbf{G}), tumor-specific T cells (\textbf{H}), and tumor-associated macrophages (\textbf{I}) as functions of $\kappa$ and $\gamma$. Colors indicate the steady-state cell populations, with blue and red denoting low and high values, respectively. The light-blue region denotes the bistable parameter regime.} 
	\label{Fig8}
\end{figure}

Figures~\ref{Fig8}A-C show the steady-state populations of tumor cells, tumor-specific T cells, and tumor-associated macrophages, respectively, as functions of $\kappa$, for several fixed values of $\gamma$. For $\gamma = 0.35$, the system converges to a stable high-tumor-burden state when $\kappa < 0.0926$ (Fig.~\ref{Fig8}A). This state corresponds to a cold tumor microenvironment, characterized by a low tumor-specific T cell population (Fig.~\ref{Fig8}B) and a high tumor-associated macrophage population (Fig.~\ref{Fig8}C). In contrast, when $\kappa > 0.1198$, the system converges to a stable low-tumor-burden state (Fig.~\ref{Fig8}A). This state displays a hot tumor phenotype, with a high tumor-specific T cell population (Fig.~\ref{Fig8}B) and a low tumor-associated macrophage population (Fig.~\ref{Fig8}C). 

Comparison of the bifurcation curves at different values of $\gamma$ further showed that increasing $\gamma$ increased the steady-state burden and tumor-associated macrophage population, while decreasing the tumor-specific T cell population (Fig.~\ref{Fig8}A-C). Thus, enhanced macrophage polarization promotes an immunosuppressive microenvironment, weakens T cell-mediated antitumor activity, and shifts the system toward an immune escape state.

For $\gamma = 0.35$, the system exhibited three positive steady states with $0.0926<\kappa<0.1198$ (Fig.~\ref{Fig8}A-C). These consisted of two stable equilibria, corresponding to high- and low-tumor-burden states, separated by an unstable equilibrium. This structure generates a bistable switch, such that the long-term outcome depends on the initial conditions (Fig.~\ref{Fig8}D-F). For example, at $\kappa = 0.1$ and $\gamma = 0.35$, the system converged to the low-tumor-burden state when the initial tumor cell and tumor-associated macrophage populations were sufficiently low (Fig.~\ref{Fig8}D-F). 

These results show that, when T cell activation is insufficient to unambiguously drive tumor control, the initial composition of the tumor microenvironment becomes a critical determinant of the eventual tumor state. Consequently, conditions during early tumor development or following treatment may determine whether the system approaches an immune-controlled or immune-escape state. This bistable behavior also provides a dynamical rationale for combination strategies that both reduce tumor burden and remodel the immunosuppressive microenvironment.

Figures~\ref{Fig8}G-I summarize the joint effect of $\kappa$ and $\gamma$ on the steady-state populations of tumor cells, tumor-specific T cells, and tumor-associated macrophages, respectively. High $\kappa$ together with low $\gamma$ produced a hot tumor phenotype, characterized by low tumor burden, a high tumor-specific T cell population, and a low tumor-associated macrophage population (Fig.~\ref{Fig8}G-I). In contrast, low $\kappa$ and high $\gamma$ produced a cold tumor phenotype with high tumor burden, a low tumor-specific T cell population, and a high tumor-associated macrophage population.

The light-blue region in Fig.~\ref{Fig8}G-I identifies parameter combinations for which the system is bistable. This region occurs when strong T cell activation is counterbalanced by substantial macrophage-mediated immunosuppression. Under these conditions, the system can support both a low-tumor-burden, immune-controlled state and a high-tumor-burden, immune-escape state, depending on the initial conditions.

\section{Discussion}

In this study, we developed a mechanism-based mathematical model for interactions among tumor cells, tumor-specific T cells, and tumor-associated macrophages. The model incorporates an arginine-mediated pathway through which tumor-associated macrophage polarization suppresses T cell activity and thereby affects tumor control. Mathematical analysis established fundamental properties of the model, including positivity and boundedness of solutions, as well as conditions related to the existence and stability of equilibria. Numerical simulations and virtual-cohort analyses were then used to examine how these interactions may contribute to tumor immune escape and heterogeneous responses to treatment.

The model reproduced the observed tumor growth patterns in immunodeficient mice, immunocompetent mice, and mice receiving arginine injection. These comparisons support the model’s capacity to capture key biological mechanisms, although they do not uniquely identify all model parameters or mechanisms. The simulations suggest that tumor-driven arginine metabolism promotes macrophage polarization and that polarized tumor-associated macrophages suppress T cell function. This feedback can favor tumor progression. The global sensitivity analysis further indicated that tumor proliferation and T cell activation have stronger effects during the earlier stages of the simulations, whereas macrophage-mediated immunosuppression becomes more influential later. Together, these results provide a dynamical interpretation of the transition from immune surveillance to immune escape within the model.

To examine interindividual heterogeneity, we developed the TVInfABC algorithm and constructed a virtual cohort of 532 parameter sets that were consistent with the available data and model assumptions. The two temporal quantities, tumor activation time $\Phi$ and T cell peak time $\Psi$, were introduced to characterize the timing of tumor growth and immune response. Within the virtual cohort, the two quantities were strongly correlated, consistent with the antigen-driven T cell activation mechanism represented in the model. In addition, the steady-state tumor-associated macrophage population was positively associated with tumor burden and negatively associated with the steady-state T cell population. These associations support the role of tumor-associated macrophages in the model as an important mediator of the transition between tumor states with relatively active and suppressed immune responses.

The virtual-cohort simulations were also used to compare 15 two-drug strategies. Across the examined parameter sets, strategies that included inhibition of arginine-mediated macrophage polarization often produced substantial tumor reduction. The preferred combination, however, depended on the parameter set. In particular, samples with a higher tumor proliferation rate were more likely to benefit from the combination of anti-ARG1 and anti-PD-1/PD-L1 therapy, especially when the baseline level of na\"{i}ve T cells was high. The dose-response analysis further showed positive Bliss interaction indices for this combination over part of the examined dose range. These results suggest that simultaneous reduction of macrophage-mediated immunosuppression and enhancement of T cell activation may be useful under suitable model conditions. They should be interpreted as model-based hypotheses for experimental evaluation rather than as direct treatment recommendations.

The bifurcation analysis provides a complementary explanation of the nonlinear response structure of the model. For intermediate values of the T cell activation rate, the model admits two stable positive steady states separated by an unstable equilibrium. Thus, for the same parameter values, distinct initial conditions can lead to either a high-tumor-burden or a low-tumor-burden state. The two-parameter bifurcation diagrams further show that this bistable regime is governed by the balance between T cell activation and macrophage polarization. In this setting, an intervention can affect the long-term tumor state not only by reducing tumor burden directly, but also by changing the relative strength of immune activation and immunosuppression. This observation provides a dynamical basis for considering combination strategies that target both burden and the immune microenvironment.

Several limitations of the present study should be noted. First, the resting macrophage population is treated as a constant parameter. This assumes rapid replenishment through the recruitment of circulating monocytes, with replenishment sufficient to balance depletion through polarization into the activated state. This approximation does not capture macrophage turnover explicitly. In vivo, the resting macrophage population is affected by both recruitment and activation-induced depletion, and these processes may vary during tumor progression and across individuals. Second, the virtual cohort was generated from the available biological data under the specified model structure. Its predicted treatment responses therefore require validation using independent experimental data. Third, the model represents several biological processes through aggregate variables and phenomenological functional forms. Extensions that distinguish macrophage subpopulations, incorporate additional immune-cell types, or include spatial heterogeneity may provide a more detailed description of the tumor microenvironment.

Future work can proceed in several directions. First, the model can be extended by allowing the resting macrophage population to evolve dynamically. This would incorporate resting macrophages as a dynamic state variable, thereby providing a more accurate representation of macrophage dynamics in the tumor microenvironment. Second, longitudinal data from perturbation experiments could be used to improve parameter identifiability and to test the predicted bistable regimes and combination effects. Third, integrating the model with multi-omics data, including single-cell sequencing and spatial transcriptomics, may allow more individualized parameterization and prediction. Such extensions would move the framework toward data-informed, patient-specific dynamical models, while requiring careful validation against independent data.

\section{Acknowledgments} We sincerely thank the reviewers for their constructive and valuable comments on the manuscript. We thank Dr. Xinjian Wang of Jiangsu University for valuable comments on this study.

\bibliographystyle{siamplain}
\bibliography{references}

\end{document}